\newcommand{\fNLl} {f_{\rm NL}^{\rm local}}

\newcommand{\LL}{{\mathrm{L}}}
\newcommand{\NL}{{\mathrm{NL}}}

\documentclass[prd,twocolumn,nofootinbib,superscriptaddress,showpacs]{revtex4}

\usepackage{graphicx,amsmath,amsfonts,amssymb,revsymb,dcolumn,epsfig,bm}

\begin{document}
\title{Mapping the large-angle deviation from Gaussianity in simulated CMB maps}

\author{A. Bernui}
\affiliation{Instituto de Ci\^encias Exatas, Universidade Federal de Itajub\'a,
37500-903 Itajub\'a -- MG, Brazil}

\author{M.J. Rebou\c{c}as}
\affiliation{Centro Brasileiro de Pesquisas F\'{\i}sicas,
Rua Dr.\ Xavier Sigaud 150,
22290-180 Rio de Janeiro -- RJ, Brazil}

\date{\today}

\begin{abstract}
The detection of the type and level of primordial non-Gaussianity in the CMB data
is essential  to probe the physics of the early universe.
Since one does not expect that a single statistical estimator can be sensitive to all
possible forms of non-Gaussianity which may be present in the data, it is important
to employ different statistical indicators to study non-Gaussianity of CMB.
In recent works we have proposed two new large-angle non-Gaussianity indicators
based on skewness and kurtosis of patches of CMB sky sphere,
and used them to find significant deviation from Gaussianity
in frequency bands and foreground-reduced CMB maps.
Simulated CMB maps with assigned type and amplitude of primordial
non-Gaussianity are important tools to determine the strength,
sensitivity and limitations of non-Gaussian estimators.
Here we investigate whether and to what extent our non-Gaussian
indicators have sensitivity to detect non-Gaussianity of local type,
particularly with an amplitude within the 7 yr Wilkinson Microwave Anisotropy Probe
(WMAP) bounds.
We make a systematic study by employing our statistical tools to generate maps
of skewness and kurtosis from several thousands of simulated maps equipped
with non-Gaussianity of local type of various amplitudes.
We show that our indicators can be used to detect large-angle local-type
non-Gaussianity only for relatively large values of the nonlinear parameter
$f_{\rm NL}^{\rm local}$.
Thus, our indicators do not have enough sensitivity to detect deviation from
Gaussianity with the nonlinear parameter within the 7 yr WMAP bounds.
This result along with the outcomes of frequency bands and foreground-reduced
analyses, suggests that non-Gaussianity captured in the previous works by our
indicators is not of primordial origin, although it might have a primordial component.
We have also made a comparative study of non-Gaussianity of
simulated maps and of the full-sky WMAP foreground-reduced
internal linear combination (ILC)-7 yr map.
An outcome of this analysis is that the level of non-Gaussianity of the ILC-7 yr map is
higher than that of the simulated maps for $f_{\rm NL}^{\rm local}$ within WMAP bounds.
This provides quantitative indications on the suitability of the
ILC-7 yr map as Gaussian reconstruction of the full-sky CMB.
\end{abstract}


\maketitle

\section{Introduction} \label{Sec.:Intro}

The statistical properties of the temperature anisotropies of CMB
radiation and of the large-scale structure of the Universe offer a powerful probe of
the physics of the early universe%
~\cite{Komatsu-2010,PNG-rev-Liguori,Xingang-Chen-2010,NG-LSS-rev}.
Thus, for example,  although a convincing detection of a significant level 
of primordial non-Gaussianity  of local type ($f_{\rm NL}^{\rm local}\gg 1$) in
CMB data would not rule out all inflationary models, it would exclude  the entire class of
single scalar field models regardless of the form of kinetic term, potential,
or initial vacuum state~\cite{CreminelliZaldarriaga2004,Komatsu-2010}.
Such a detection, however, would be consistent with some alternative models of 
the physics of the early universe, and also with other nonstandard inflationary scenarios.
On the other hand, a convincing null detection  of primordial non-Gaussianity
of local type ($f_{\rm NL}^{\rm local} \simeq 0 $) would rule out the current
alternative models of the primordial universe
(see, e.g., Refs.~\cite{Komatsu-2010,PNG-rev-Liguori,Xingang-Chen-2010,
NG-Alternative-Models}).
Thus, a detection or null detection of primordial non-Gaussianity of local type
in the CMB data is crucial not only to discriminate or even exclude classes of
inflationary models, but also to test alternative scenarios, offering therefore
an important window into the physics of the early universe.

However,  there are various  effects that can produce non-Gaussianity.
Among them, the  most significant are possibly unsubtracted diffuse
foreground emission \cite{Bennett-etal-2003,Leach-etal-2008},
unresolved point sources~\cite{Komatsu-etal-2003}, possible systematic
errors~\cite{Su-Yadav-etal2010}, and secondary anisotropies such as
gravitational weak lensing and the Sunyaev-Zeldovich
effect (see, e.g., Refs.~\cite{Komatsu-2010,PNG-rev-Liguori}).%
\footnote{For most of these effects there seems to be no evidence of
significant non-Gaussian contamination within Wilkinson Microwave Anisotropy Probe
(WMAP) sensitivity. Nevertheless,
this is not guaranteed to hold true for the PLANCK experiment~\cite{Planck-Collab},
due to its much higher sensitivity~\cite{PNG-rev-Liguori}.
Cosmic variance further complicates the problem, since some non-Gaussianity
may arise from the uniqueness of the observed CMB sky. Deviation from 
Gaussianity may also have a cosmic topology origin
Refs.~\cite{CosmTopReviews}.}
In this way, the accurate extraction of a possible primordial non-Gaussianity is
a challenging observational and statistical enterprise.

In the search for non-Gaussianity in CMB data, different statistical tools
can, in principle, provide information about distinct forms of non-Gaussianity.
On the other hand, one does not expect that a single statistical estimator
can be sensitive to all possible forms of non-Gaussianity that may be present
in CMB data.
It is therefore important to test CMB data for deviations from
Gaussianity by using  different statistical tools in order to
quantify or constrain the amount of any non-Gaussian signals in the
data, and extract information on their possible origins.
This has motivated a great deal of effort that has recently gone into
the search for non-Gaussianity in CMB maps by employing several statistical
estimators (an incomplete list of references is given the recent
reviews~\cite{Komatsu-2010,PNG-rev-Liguori,Xingang-Chen-2010}, and also in
Refs.~\cite{Bernui-Reboucas2009,Bernui-Reboucas2010,Bernui-etal2007}) 

In recent papers~\cite{Bernui-Reboucas2009,Bernui-Reboucas2010}
we have proposed two new large-angle non-Gaussianity estimators,
which are based upon the skewness and kurtosis of the
patches (spherical caps) of the CMB sky sphere.
By scanning  the CMB sphere with evenly distributed
spherical caps, and calculating the skewness $S$ and kurtosis $K$ for
each cap, one can have measures of the departure from Gaussianity
on large angular scales.
Using this procedure we have carried  out analyses of large-angle
deviation from Gaussianity in both band and foreground-reduced maps with and
without the  \emph{KQ75} mask recommended by the Wilkinson Microwave Anisotropy
Probe (WMAP) team for Gaussianity studies of CMB data%
~\cite{Bernui-Reboucas2009,Bernui-Reboucas2010}.
Thus, we have found, for example, significant departure from Gaussianity in
all full-sky band maps, while for the Q, V, W  maps with the \emph{KQ75} mask,
the CMB data are consistent with Gaussianity.
The K and Ka maps, however, show an important deviation from Gaussianity even
with a \emph{KQ75} mask.
For the full-sky foreground-reduced%
~\cite{ILC-5yr-Hishaw,ILC-7yr-Gold,HILC-Kim,NILC-Delabrouille}
we have found  a significant deviation from Gaussianity
%
which varies with the cleaning processes, as measured by our
indicators~\cite{Bernui-Reboucas2010}.

Simulated CMB maps equipped with assigned primordial non-Gaussianity
are essential tools to test the sensitivity and effectiveness of non-Gaussian
indicators. They can also be used to study, e.g., the effects of foregrounds
and other non-Gaussian contaminants, and also to disclose potential systematics.
In a recent paper a new algorithm for generating non-Gaussian CMB temperature maps
with non-Gaussianity of the local type has been devised~\cite{ElsnerWandelt2009}.
In the simulated maps the level of non-Gaussianity is adjusted by the
dimensionless parameter $f_{\rm NL}^{\rm local}$. A set of $1\,000$ CMB temperature
simulated maps with the resolution of PLANCK mission~\cite{Planck-Collab} for open
(unfixed) values of $f_{\rm NL}^{\rm local}$ was made available~\cite{ElsnerWandelt2009}.%
%

A pertinent question that arises is whether the indicators proposed in 
Ref.~\cite{Bernui-Reboucas2009}, and used in
Refs.~\cite{Bernui-Reboucas2009,Bernui-Reboucas2010} to detect non-Gaussianity in
CMB data, have sufficient sensitivity to detect deviation from non-Gaussianity of local
type with an amplitude $f_{\rm NL}^{\rm local}$ within the bounds determined by the
WMAP team in their latest data release~\cite{WMAP7}.
Our primary aim in this paper is to address this question  by extending
the results, and by complementing the analyses of
Refs.~\cite{Bernui-Reboucas2009,Bernui-Reboucas2010} in  three different ways.
First, instead of using CMB data, we use our statistical indicators to
carry out an analysis of Gaussianity of \emph{simulated} maps equipped with
non-Gaussianity of local type.
Second, by using simulated maps with different amplitudes $f_{\rm NL}^{\rm local}$,
we make a quantitative analysis of the degree of sensitivity of our skewness and
kurtosis indicators to detect primordial non-Gaussianity of local type. 
Third, we make a comparative study of the degrees of non-Gaussianity of simulated maps
with different $f_{\rm NL}^{\rm local}$'s
and the full-sky foreground-reduced 7 yr WMAP internal linear
combination (ILC-7 yr) map~\cite{ILC-7yr-Gold}.
An interesting outcome of this comparative analysis is that the level of non-Gaussianity
of the ILC-7 yr is considerably higher than that of the simulated maps for
$f_{\rm NL}^{\rm local}$ within observational bounds~\cite{WMAP7}.
This renders information about the suitability of the ILC-7 yr foreground-reduced map
as Gaussian reconstruction of the CMB full-sky.

\section{Non-Gaussianity and simulated maps}

\subsection{Primordial non-Gaussianity of local type} \label{Sec:NG-Local}

The first important ingredient in the study of non-Gaussianity is the
primordial gravitational curvature perturbations $\zeta(\mathbf{x},t)$,
which were seeded  by quantum fluctuations in the very early universe.
In the linear order, the primordial curvature perturbation is related
to Bardeen's curvature perturbation $\Phi(\mathbf{x},t)$~\cite{Bardeen1980}
in the matter dominated era by $\zeta= (5/3)\,\Phi$~\cite{Kodama1984}.
The relation with CMB observations is given in the Sachs-Wolfe limit,
where $\Delta T/T = - \,\Phi/3 = - \zeta / 5$ holds~\cite{SachWolfe1967}.

The lower order statistics able to distinguish non-Gaussian from Gaussian
distributions is the three-point correlation function.
Primordial non-Gaussianity can  then be described in terms of
the three-point correlation function of the curvature perturbations
$\Phi(\mathbf{x})$ or of its Fourier transform $\Phi(\mathbf{k})$
by using the ensemble average.
Thus, the three-point correlation function counterpart in Fourier space
---the so-called bispectrum--- takes the form
\begin{equation}
\langle \Phi(\mathbf{k_1}) \Phi(\mathbf{k_2}) \Phi(\mathbf{k_3})
\rangle = \delta^{3}(\mathbf{k_{123}})\,B^{}_\Phi(k_1,k_2,k_3) \;, 
\end{equation}
where $\delta^{(3)} (\mathbf{k_{123}})\equiv \delta^{(3)}(\mathbf{k_1 + k_2 + k_3})$
enforces that the wavevectors in Fourier space have to close to form a triangle, i.e.,
$\mathbf{k_1 + k_2 + k_3}=0$.
Thus, the bispectrum is a function of the triplet defined by the magnitude of the
wave numbers $(k_1, k_2,k_3)$.

In the examination of primordial non-Gaussianity, the bispectrum of curvature
is rewritten in the form
\begin{eqnarray} 
B^{}_\Phi (\mathbf{k_1}, \mathbf{k_2},\mathbf{k_3}) &=& \langle \Phi(\mathbf{k_1})
\Phi(\mathbf{k_2}) \Phi(\mathbf{k_3}) \rangle                       \nonumber \\
&=& \!\!f^{}_{\rm NL} \,(2\pi)^3\, \delta^{3} (\mathbf{k_{123}}) \, F (k_1,k_2,k_3),
\end{eqnarray}
where $f^{}_{\rm NL}$ is an overall amplitude (dimensionless) parameter,%
\footnote{The subscript 'NL' stands for nonlinear. This notation arises
because an often used phenomenological parametrization for the
non-Gaussianity of $\Phi(\mathbf{x})$ can be written as a nonlinear
transformation of a Gaussian field (see below for more details).}
which can be constrained by the CMB data, and where $F (k_1,k_2,k_3)$ is the
so-called shape of the bispectrum and encodes the functional dependence of the
bispectrum on the triangular configurations.
The bispectrum shape is used to specify the configuration of the wave vectors
of curvature perturbations that give the highest contributions to the
amplitude of the bispectrum.
In this way, the bispectra are usually classified according to shapes
of the triangles that give rise to the highest non-Gaussian signal.
The most studied shapes are  {\bf (i)} \emph{local}, where
the bispectrum is maximum on the squeeze triangle configuration,
$k_1 \approx k_2 \gg k_3$; {\bf (ii)} \emph{equilateral}, with the bispectrum
maximized on the equilateral limit $k_1 \approx k_2 \approx k_3$;
and {\bf (iii)} \emph{orthogonal}, where
the bispectrum has a positive peak at the equilateral configuration and
a negative valley along the flattened (elongated) triangular configuration
$k_3 \approx k_1+k_2$~\cite{Komatsu-2010}.
Predictions for both the amplitude $f^{}_{\rm NL}$ and the shape of
$B^{}_\Phi (k_1, k_2,k_3)$ depend on the early universe models,
which makes apparent the power of the bispectrum as a tool to probe
models of the primordial universe.

In this paper we shall deal with non-Gaussianity of local type, which is
the most studied type of deviation from Gaussianity.
For this type of primordial non-Gaussianity the curvature perturbation in the real
space can be split into two components: the linear term $\Phi_\LL$ (representing
the Gaussian component) plus a non-Gaussian second term, namely%
\footnote{It should be note, however, that this is not the only way to produce
the  bispectrum of local type. Multi-field curvaton inflation, for example, can
also produce bispectrum of local type~\cite{Lyth-etal_2003}.}
\begin{equation} \label{local_param}
\Phi (\mathbf{x})=\Phi_\LL (\,\mathbf{x})+f_{\rm NL}^{\rm local}\,( \,\Phi_L^2(\mathbf{x})
                            - \langle \,\Phi_\LL^2(\mathbf{x}) \,\rangle\,)\;,
\end{equation}
where both sides are evaluated at the same spatial location $\,\mathbf{x}$.
%

Finally, we recall that the latest WMAP constraints reported on $ f_{\rm NL}^{\rm local}$
at the $95\%$ confidence level are given by
$ -10 <  f_{\rm NL}^{\rm local} < 74$~\cite{WMAP7}. These
bounds will be used in our analyses in one of the following sections.

\subsection{Simulated CMB maps} \label{Sec:Sim-Maps}

The first simulated CMB temperature maps endowed with primordial
non-Gaussianity introduced through a non-Gaussian parameter $f_{\rm NL}$
were generated  by the WMAP team and reported in Ref.~\cite{Komatsu-etal-2003}.
The WMAP algorithm was generalized so as to improve computational speed
and accuracy, and also to include polarization maps in  
Ref.~\cite{Liguori_2003} (see also Ref.~\cite{Liguori_2007}).
More recently, Elsner and Wandelt~\cite{ElsnerWandelt2009} presented a new
algorithm and generated $1\,000$ high-angular resolution simulated
non-Gaussian CMB temperature and polarization maps with non-Gaussianities of
the local type, for which the level of non-Gaussianity is determined by the
parameter $f_{\rm NL}^{\rm local}$.
In their algorithm a simulated map with a desired level of non-Gaussianity
$f_{\rm NL}^{\rm local}$ is such that the spherical harmonic coefficients
are given by
\begin{equation}
a_{\ell m} = a^\LL_{ \ \, \ell m} + \fNLl \cdot a^\NL_{\ \ \,\,\ell m}\,,
\end{equation}
where $a^\LL_{ \ \, \ell m}$ and $a^\NL_{\ \ \,\,\ell m}$ are, respectively,
the linear and nonlinear spherical harmonic coefficients of the simulated CMB
temperature maps.

In the next sections  we shall use the set of $6,000$ CMB temperature simulated linear
and nonlinear component maps, generated for an arbitrary (unfixed) value of
$f_{\rm NL}^{\rm local}$, which was made available in Ref.~\cite{ElsnerWandelt2009}%
\footnote{
{\tt http://planck.mpa-garching.mpg.de/cmb/fnl-simulations}.}
with the resolution of the PLANCK mission~\cite{Planck-Collab}.

\section{Non-Gaussianity Indicators and associated maps} \label{Sec:Indicators}

\begin{figure*}[t!] 
\begin{center}
\includegraphics[width=4.5cm,height=6.5cm,angle=90]{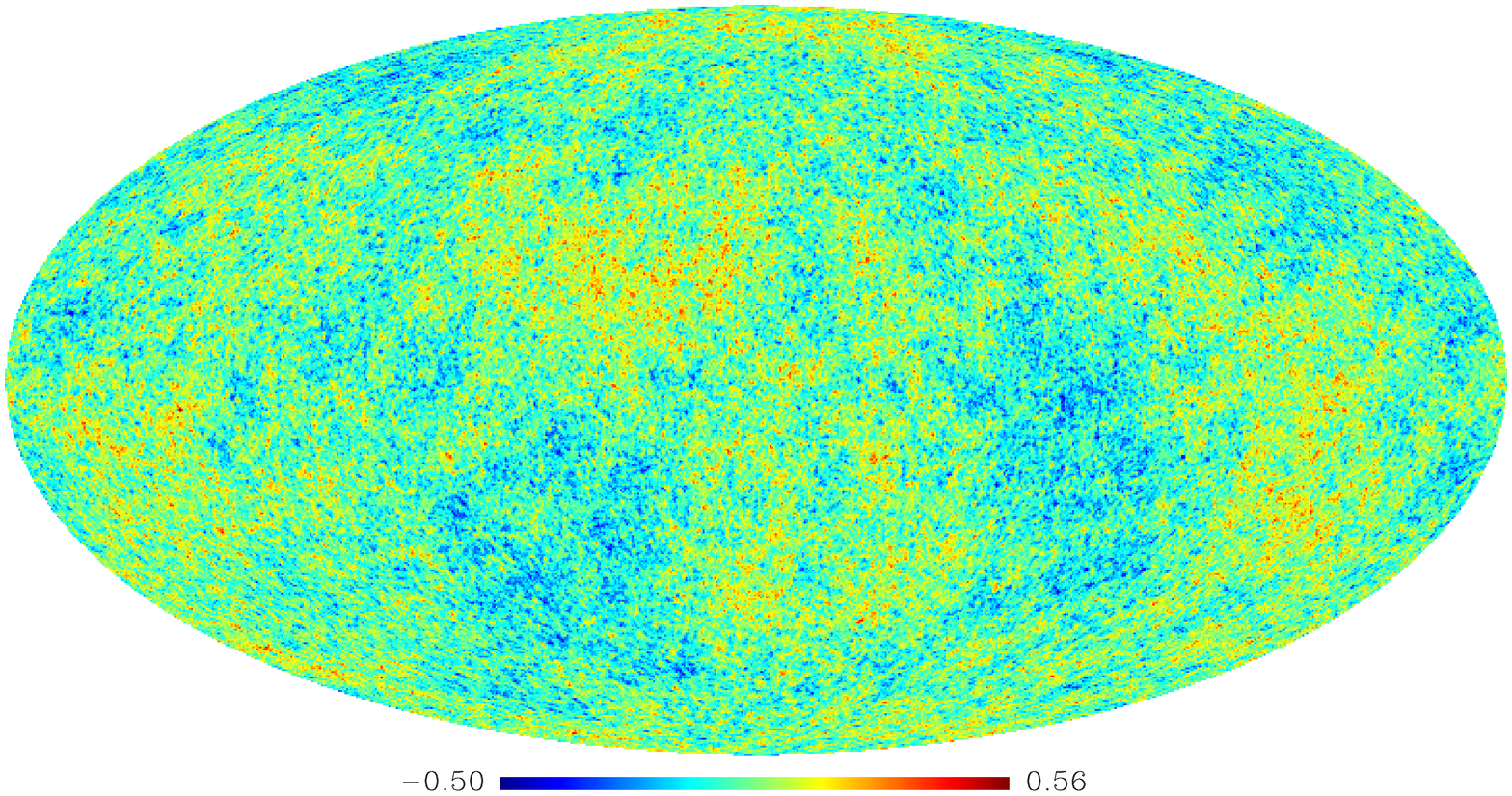}
\hspace{10mm}
\includegraphics[width=4.5cm,height=6.5cm,angle=90]{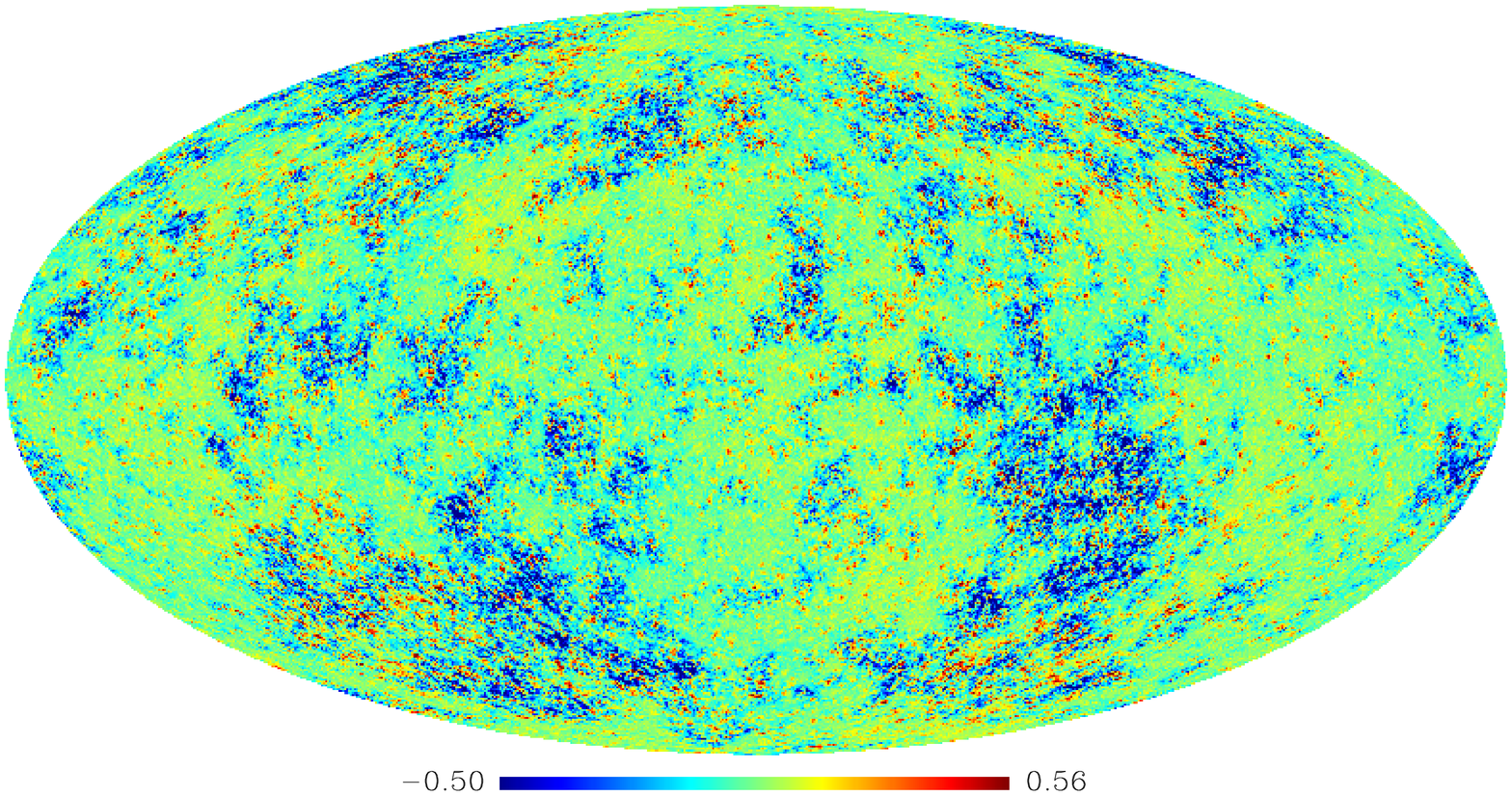}
\caption{\label{Fig1}  The left panel shows a simulated Gaussian CMB map
($f_{\rm NL}^{\rm local}=0$), while the right panel depicts a non-Gaussian simulated
CMB map calculated for $f_{\rm NL}^{\rm local}=5\,000$.  Temperatures are in $mK$.}
\end{center}
\end{figure*}
To make our paper clear and self-contained,
in this section we describe the two statistical non-Gaussianity indicators
and their associated maps, which can be calculated from either simulated or real
CMB temperature (input) maps. The procedure delineated here will be used in
the following sections to investigate large-angle deviation from Gaussianity.

The most important underlying idea in the construction of our non-Gaussianity
indicators and the associated maps, is  that one can access the deviation from
Gaussianity of the CMB temperature fluctuations by calculating  the skewness
$S$ and the kurtosis $K$, which measure, respectively, the symmetry about
the mean temperature, and the non-Gaussian degree of peakness of the
temperature distribution.

Clearly the calculation of  $S$ and $K$ from the whole CMB sky sphere
of temperature fluctuations data would be a crude approach to measure
the complexity of the non-Gaussianity of the CMB map. However, instead
of having just two dimensionless numbers, one can go further by taking a discrete
set of points $j=1, \ldots ,N_\mathrm{c}$ homogeneously distributed on the CMB
sky sphere $S^2$ as the center of \emph{spherical caps} of aperture $\gamma$ (say),
and calculate $S_j$ and $K_j$ for each cap $j\,$ of the CMB temperature sky sphere.
The values $S_j$ and $K_j$ can then be taken as measures of the
non-Gaussianity in the direction $(\theta_j, \phi_j)$ of the center
of the spherical cap $j\,$.
Such calculations for the individual caps thus provide quantitative information
($2 N_\text{c}$ values) about non-Gaussianity of the CMB data.

The above procedure is a constructive way of defining on $S^2$ two
discrete functions $S(\theta, \phi)$ and $K(\theta, \phi)$ that
measure departure from Gaussianity of a CMB temperature (input) map.
In other words, these functions  can be constructed from an
input CMB map through the following steps~\cite{Bernui-Reboucas2009}:
\begin{enumerate}
\item[{\bf i.}]
Take a discrete finite set of points $j=1, \ldots ,N_{\rm c}$ homogeneously
distributed on the sky sphere of a CMB input map as the centers of spherical
caps of a chosen aperture $\gamma$; and calculate for each cap $j$ the skewness
and kurtosis  given, respectively,  by
\begin{eqnarray} \label{S_and_K_def}
S_j  & = & \frac{1}{N_{\mbox{\footnotesize p}} \,\sigma^3_j }
\sum_{i=1}^{N_{\mbox{\footnotesize p}}} \left(\, T_i\, - \overline{T} \,\right)^3 \;,  \\
K_j  & = & \frac{1}{N_{\mbox{\footnotesize p}} \,\sigma^4_j }
\sum_{i=1}^{N_{\mbox{\footnotesize p}}} \left(\,  T_i\, - \overline{T} \,\right)^4 - 3 \;,
\end{eqnarray}
where
$T_i$ is the temperature at the $i^{\,\rm{th}}$ pixel, $\overline{T_j}$ is
the CMB mean temperature of the  $j^{\,\rm{th}}$ spherical cap,
$N_{\mbox{\footnotesize p}}$ is the number of pixels in the cap $j\,$, 
and $\sigma^2 = (1/N_{\mbox{\footnotesize p}}^{}) \sum_{i=1}^{N_{\mbox{\footnotesize p}}^{}}
\left(\, T_i\, - \overline{T} \,\right)^2$ is the standard  deviation.

Clearly, the whole set of values $S_j$ and $K_j$ (for all $j$) obtained
through this discrete \emph{scanning} calculation process, along with %
the angular coordinate of the center of the caps $(\theta_j, \phi_j)$
can be taken as measures of non-Gaussianity in the directions of the
center of spherical caps $j\,$.  
\item[{\bf ii.}]
Patching together the $S_j$ and $K_j$ values for all spherical caps $j$, one has two
discrete functions $S = S(\theta,\phi)$ and $K = K(\theta,\phi)$ defined over a
two-sphere $S^2$.
These functions provide measurements of the non-Gaussianity as a function of
$(\theta,\phi)$.
The Mollweide projections of the functions $S = S(\theta,\phi)$ and $K = K(\theta,\phi)$
are nothing but skewness and kurtosis maps (hereafter denoted by $S$ map and $K$
map).
\end{enumerate}

Clearly, the functions $S = S(\theta,\phi)$ and $K = K(\theta,\phi)$
can be expanded into their spherical harmonics. 
Thus, for example, for  $S = S(\theta,\phi)$ 
one has
\begin{equation}
S(\theta,\phi) = \sum_{\ell=0}^\infty \sum_{m=-\ell}^{\ell}
b_{\ell m} \,Y_{\ell m} (\theta,\phi) \; ,
\end{equation}
and  the corresponding angular power spectrum
\begin{equation}
S_{\ell} = \frac{1}{2\ell+1} \sum_m |b_{\ell m}|^2 \; ,
\end{equation}
which can be used to quantify the amplitude (level) and
angular scale of the deviation from Gaussianity. In this paper
we shall use the power spectra $S_\ell$ and $K_\ell$ to estimate
the departure from Gaussianity and calculate the statistical
significance of such a deviation by comparison with the
corresponding power spectra calculated from $S$ and $K$ maps
obtained from input Gaussian maps ($f_{\rm NL}^{\rm local}=0$).

In the next section we will  use these  statistical indicators to
carry out analyses of Gaussianity of \emph{simulated} maps endowed
with non-Gaussianity of local type of different levels to test the
sensitivity of $S$ and $K$ indicators to detect primordial
non-Gaussianity of local type. Furthermore, we will make a comparative
study of the degrees of non-Gaussianity of different simulated maps
and the ILC-7 yr WMAP map.

\section{Non-Gaussianity and sensitivity of $S$ and $K$ indicators}
\label{Sec:Indic_vs_data}

\subsection{Analyses and results for simulated maps} \label{SubSec:Analy-Sim-Maps}

\begin{figure*}[t] 
\begin{center}
\includegraphics[width=4.5cm,height=6.5cm,angle=90]{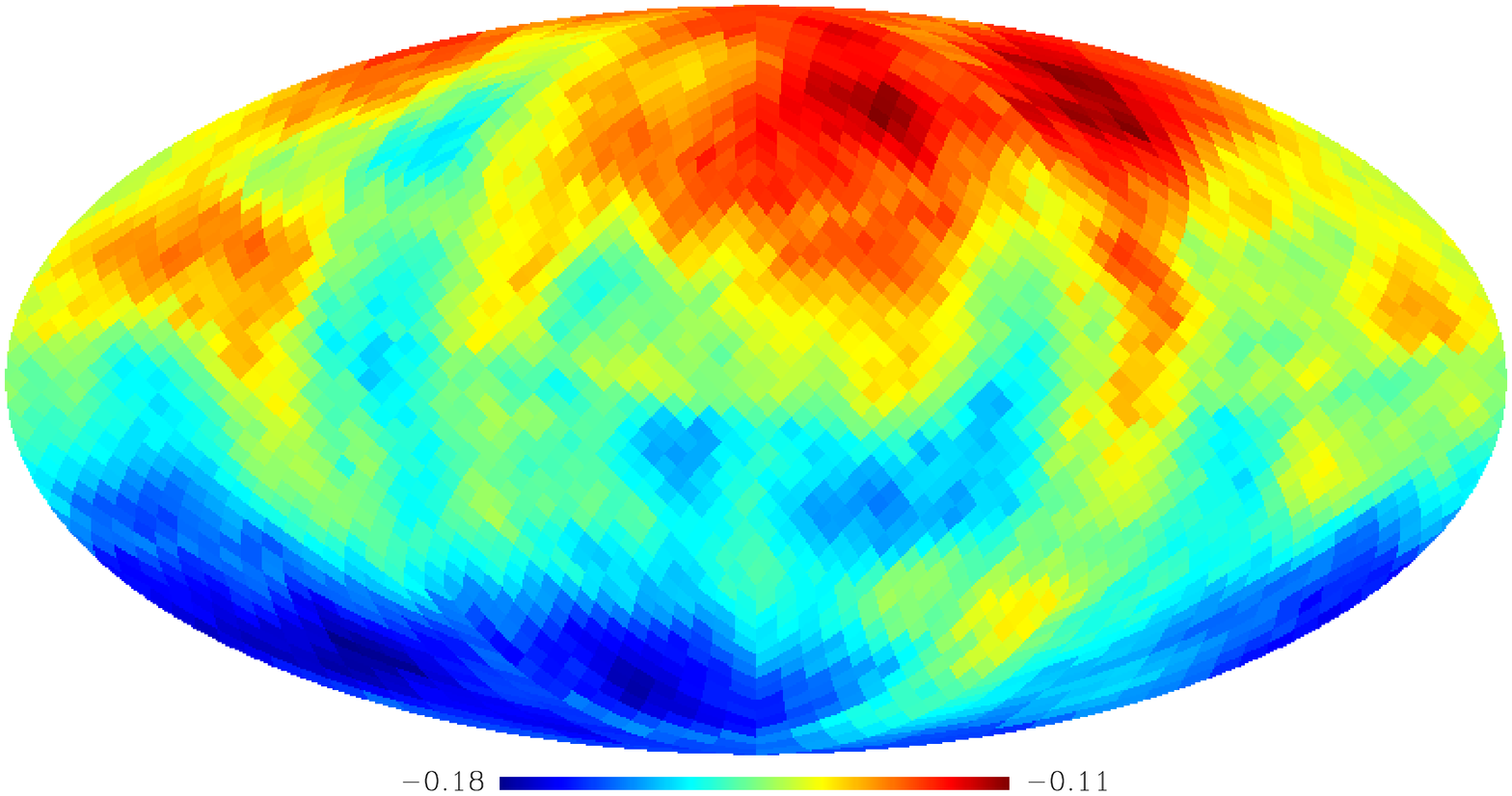}
\hspace{10mm}  
\includegraphics[width=4.5cm,height=6.5cm,angle=90]{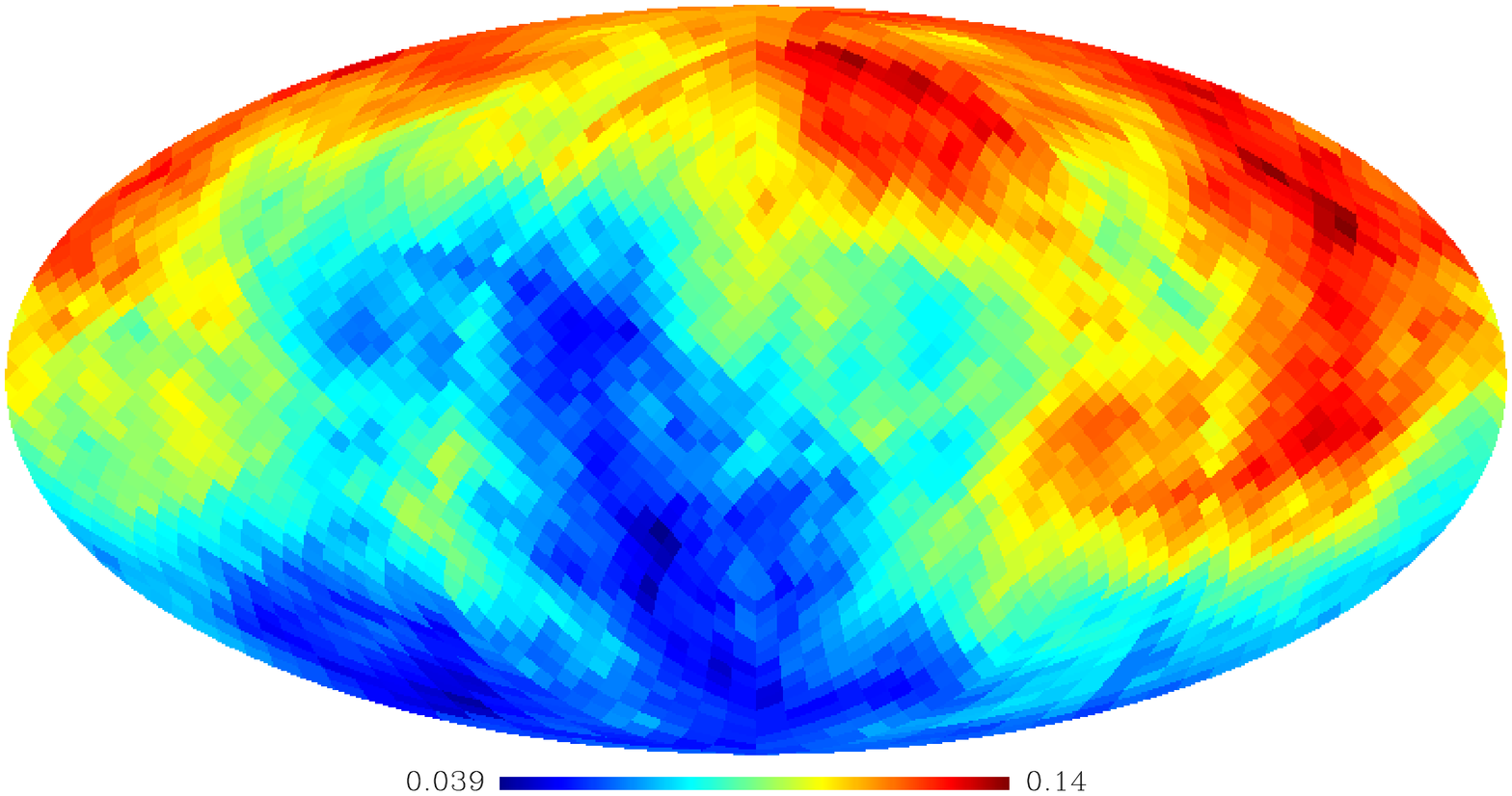}
\vskip 3mm
\includegraphics[width=4.5cm,height=6.5cm,angle=90]{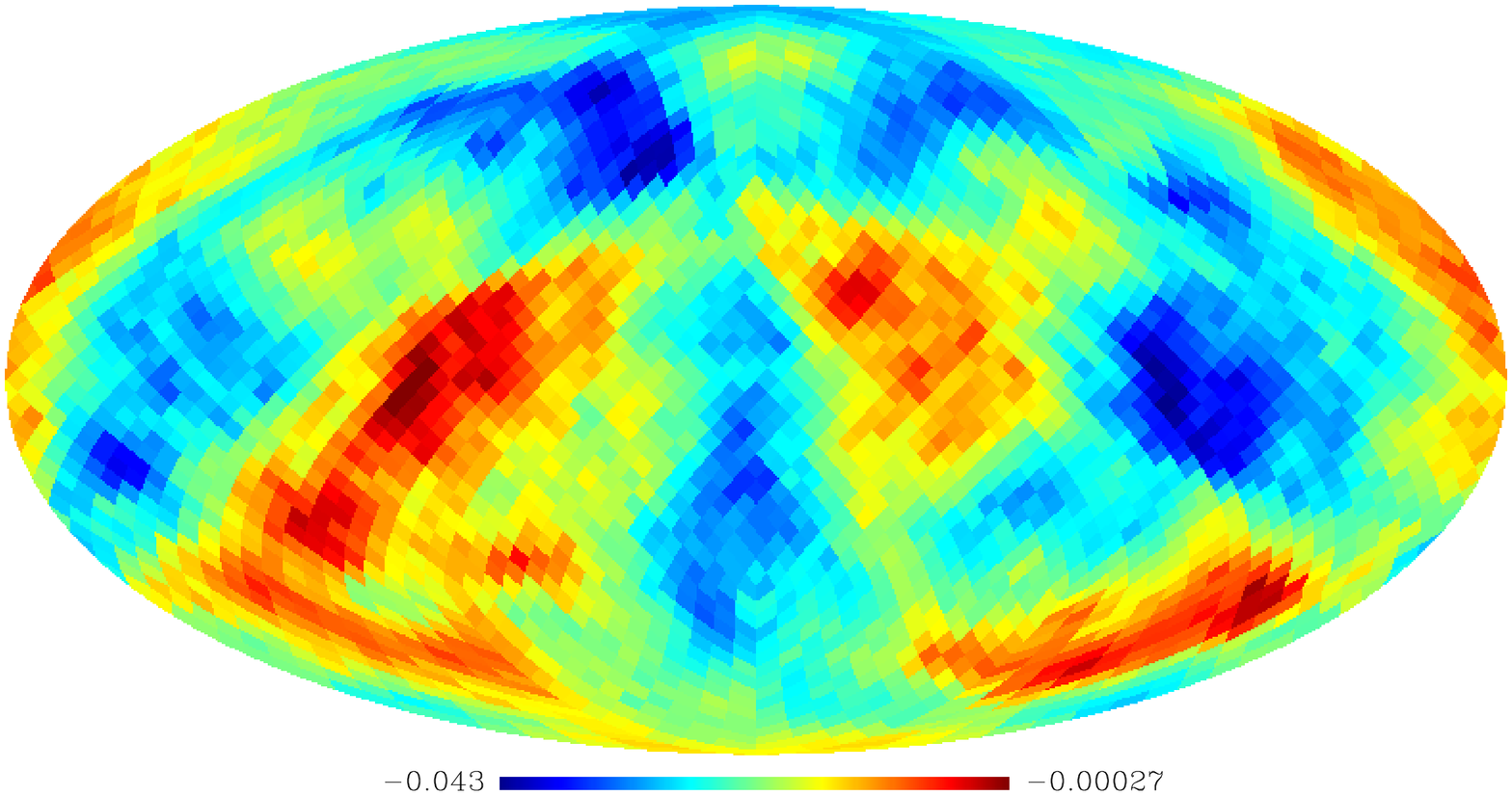}
\hspace{10mm}
\includegraphics[width=4.5cm,height=6.5cm,angle=90]{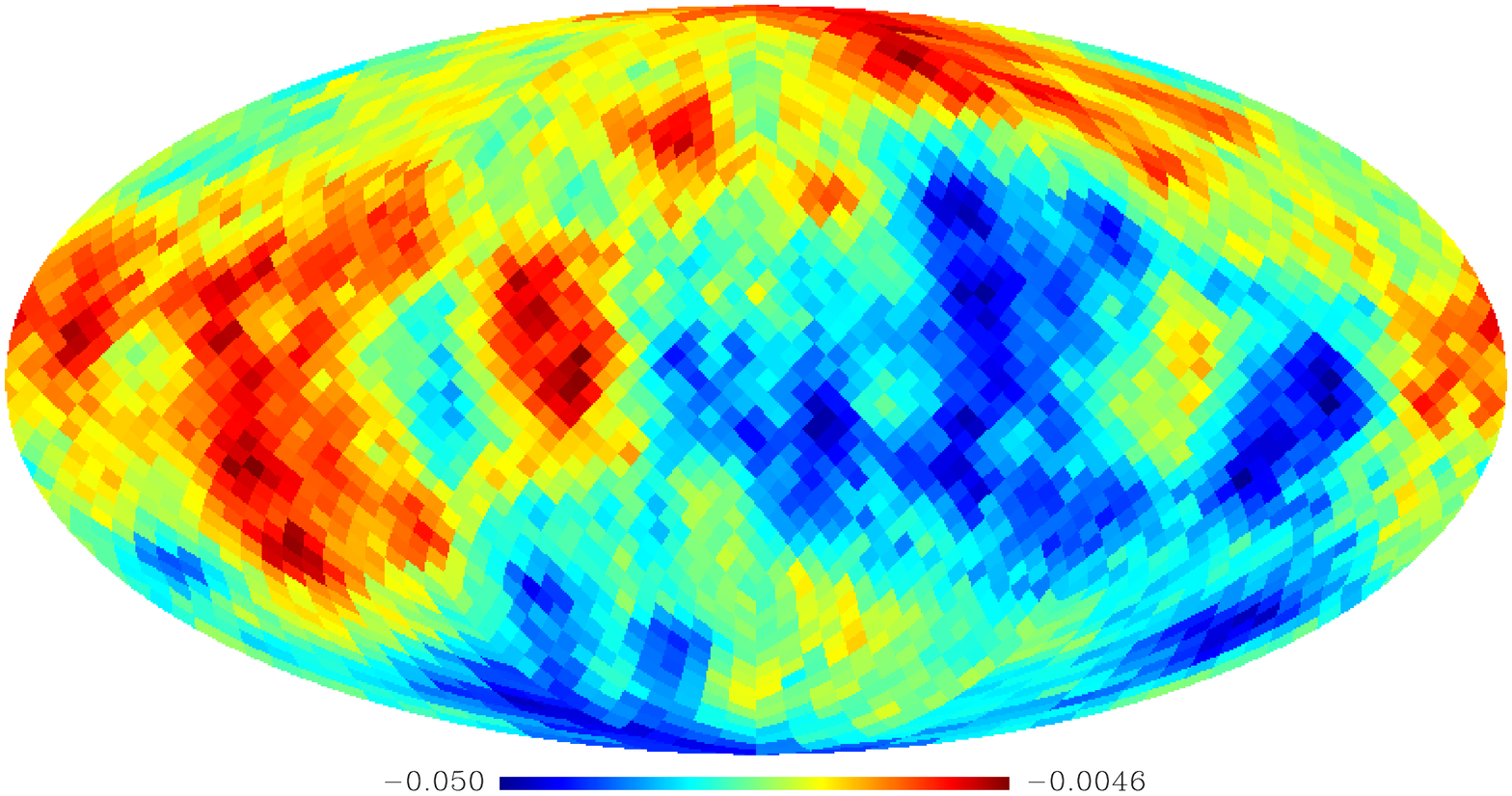}
\caption{\label{Fig2} Examples of skewness (left panels) and kurtosis (right panels)
indicator maps calculated from simulated maps with, respectively, $\fNLl = 500$
(first row) and $\fNLl = 0$
(second row).}
\end{center}
\end{figure*}

It is clear from the previous section that in order to study the sensitivity of the
non-Gaussianity indicators $S$ and $K$, that is, to construct the functions
$S = S(\theta,\phi)$ and $K = K(\theta,\phi)$ and the associated $S$ and $K$
maps, 
we need CMB input maps.
The input maps used in our analyses in this section are high-angular
resolution-simulated CMB temperature maps endowed with non-Gaussianities
of the local type introduced through different values of the dimensionless
amplitude parameter $\fNLl$.

Figure~\ref{Fig1} shows two examples of such simulated CMB maps.
The left panel gives a Gaussian map $\fNLl = 0\,$, while the right
panel shows a non-Gaussian map for $\fNLl = 5\,000\,$. We have taken these
values for $f_{\rm NL}^{\rm local}$ as an  illustrative example 
that makes the non-Gaussian effects visible to the naked eye through the
comparison of these simulated maps.

Figure~\ref{Fig2} gives an illustration of typical skewness $S$ (left panel)
and kurtosis $K$ (right panel) maps, generated from input CMB simulated
maps for $f_{\rm NL}^{\rm local}=500$ (first row) and $f_{\rm NL}^{\rm local}=0$
(second row).
The non-Gaussian maps show spots with higher and lower values of
$S(\theta,\phi)$ and $K(\theta,\phi)$, which suggest \emph{large-angle}
dominant multipole components (low $\ell$) in these maps.%
\footnote{We have also calculated similar maps from the simulated
maps for the other values of $f_{\rm NL}^{\rm local}$.
These maps, however, provide only qualitative information, and to avoid
repetition we only depict maps of Fig.~\ref{Fig2} for illustrative purpose.}
In our calculations of all $S$ maps and $K$ maps, to minimize the statistical noise,
we have scanned the CMB sphere with $N_{\rm c}=3\,072$ spherical caps of aperture
$\gamma = 90^{\circ}$, centered at points homogeneously distributed on the two-sphere.

To obtain \emph{quantitative} large-angle-scale information, as a first test of the
sensitivity of $S$ and $K$ indicators, for each of the following values of the nonlinear
parameter
--$\fNLl = 0, 500, 1\,000, 5\,000\,$-- we have generated   
$1\,000$ simulated CMB maps, totalizing $4\,000$ simulated
maps.
From each set of $1\,000$ simulated input CMB maps (fixed $\fNLl$ for each set) we
have calculated  $1\,000$ $S$ maps and $1\,000$ $K$ maps, from which we
computed the associated power spectra, namely,
$S^{\,\mathbf{i} }_{\,\ell}$ and $K^{\,\mathbf{i} }_{\,\ell}$,
where $\,\mathbf{i}= 1,\,\,\cdots,1\,000\,$ is an enumeration index,
and $\,\ell=1,\,\,\cdots,10\,$ is the range of multipoles %
we have focused on in this paper.%
\footnote{The values of  $\ell_{max}$ for $S_{\ell}$ and  $K_{\ell}$ depend
on the resolution of $S$ and $K$ maps, which clearly depend upon
the number of spherical caps used in the scanning process.
For $N_{\rm c}=3\,072$ one can go up to  $\ell_{max} = 45$.}
The low $\ell$ multipole mean components $S_{\ell}$ and $K_{\ell}$ are then obtained
by averaging over $1\,000$ power spectra calculated from simulated maps with
the different values of $f_{\rm NL}^{\rm local}$.
The statistical significance of these power spectra is estimated
by comparing the values of $S_{\ell}$ and $K_{\ell}$ obtained from 
input maps generated for  $f_{\rm NL}^{\rm local}= 500,
1\,000, 5\,000$ with the values of the corresponding power spectra
$S_{\ell}$ and $K_{\ell}$ obtained from the Gaussian
simulated map ($f_{\rm NL}^{\rm local}= 0$).   

Let us describe with some details the calculations of an
average power spectrum.
For the sake of brevity, we focus on the skewness indicator $S$
along with a set of $1\,000$  simulated non-Gaussian input maps computed
for a nonlinear parameter $\fNLl= 500$, for example.
\footnote{A completely similar procedure can clearly be used
for the kurtosis indicator $K$ along with other values of  $\fNLl$.}
Following the two step procedure described in Sec.~\ref{Sec:Indicators},
from these $1\,000$ simulated CMB maps we calculated $1\,000$ $S$ maps,
from which we computed $1\,000$ power spectra $S^{\,\mathbf{i} }_{\,\ell}$
in order to have the average values $S_{\ell} =  
(1/1000) \sum_{\mathbf{i} = 1}^{1000} S_{\,\ell}^{\,\mathbf{i}}\,$.
{}From this MC process we have at the end ten mean multipole values
$S_{\ell}\,$ ($\,\ell=1,\,\,\cdots,10\,$), each of which is then used for
a comparison with the corresponding average multipole value $S_{\ell}$
calculated by a similar procedure from Gaussian simulated maps ($\fNLl = 0$).
This allows the evaluation of the statistical significance of
$S_{\ell}$  by quantifying  
the goodness of fit for $S_{\ell}$ with $\fNLl \neq 0$ and $S_{\ell}$
calculated for $\fNLl = 0$ (Gaussian maps).

\begin{figure*}[htb!]
\begin{center}
\includegraphics[width=8.8cm,height=5.6cm]{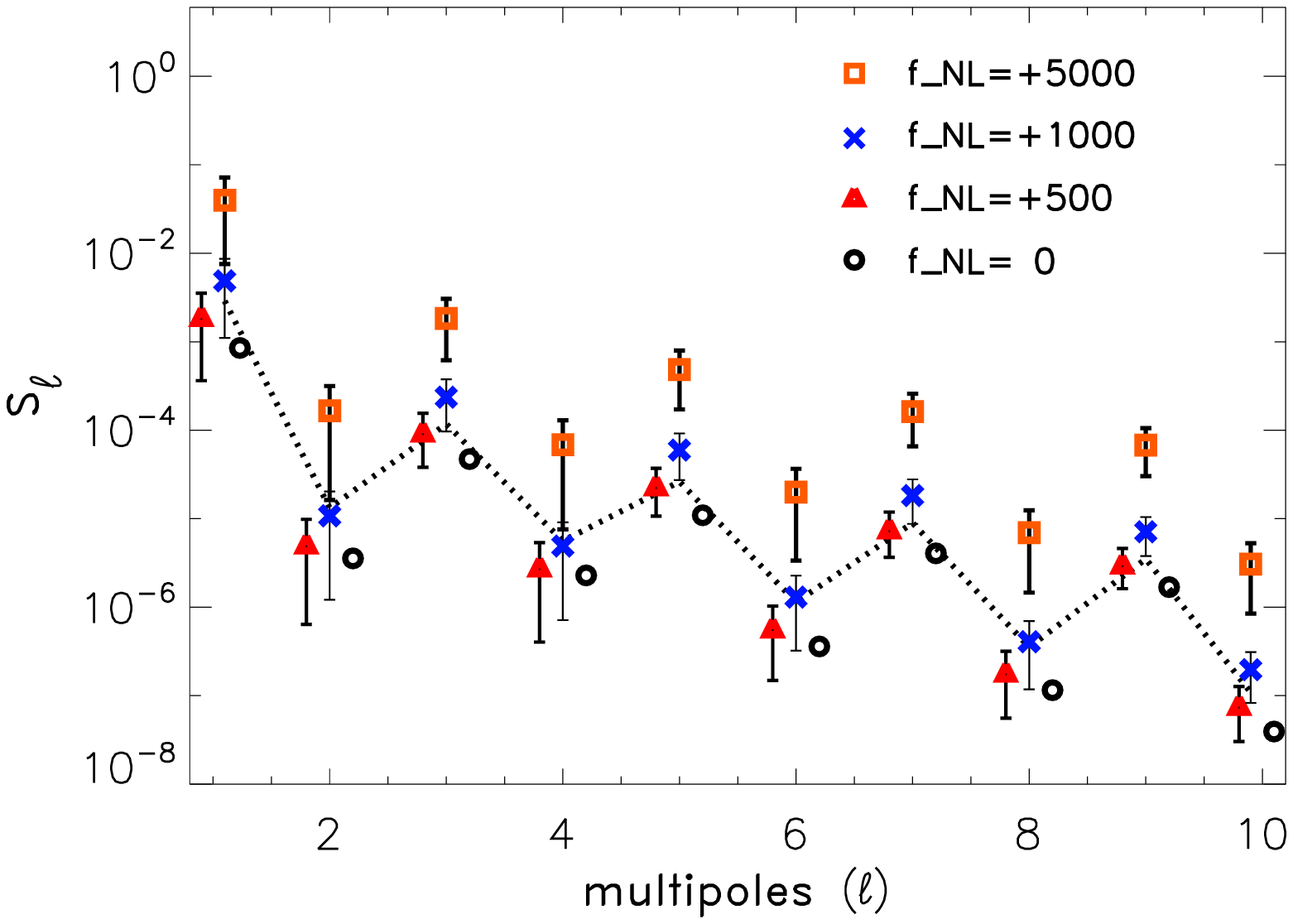}
\includegraphics[width=8.8cm,height=5.6cm]{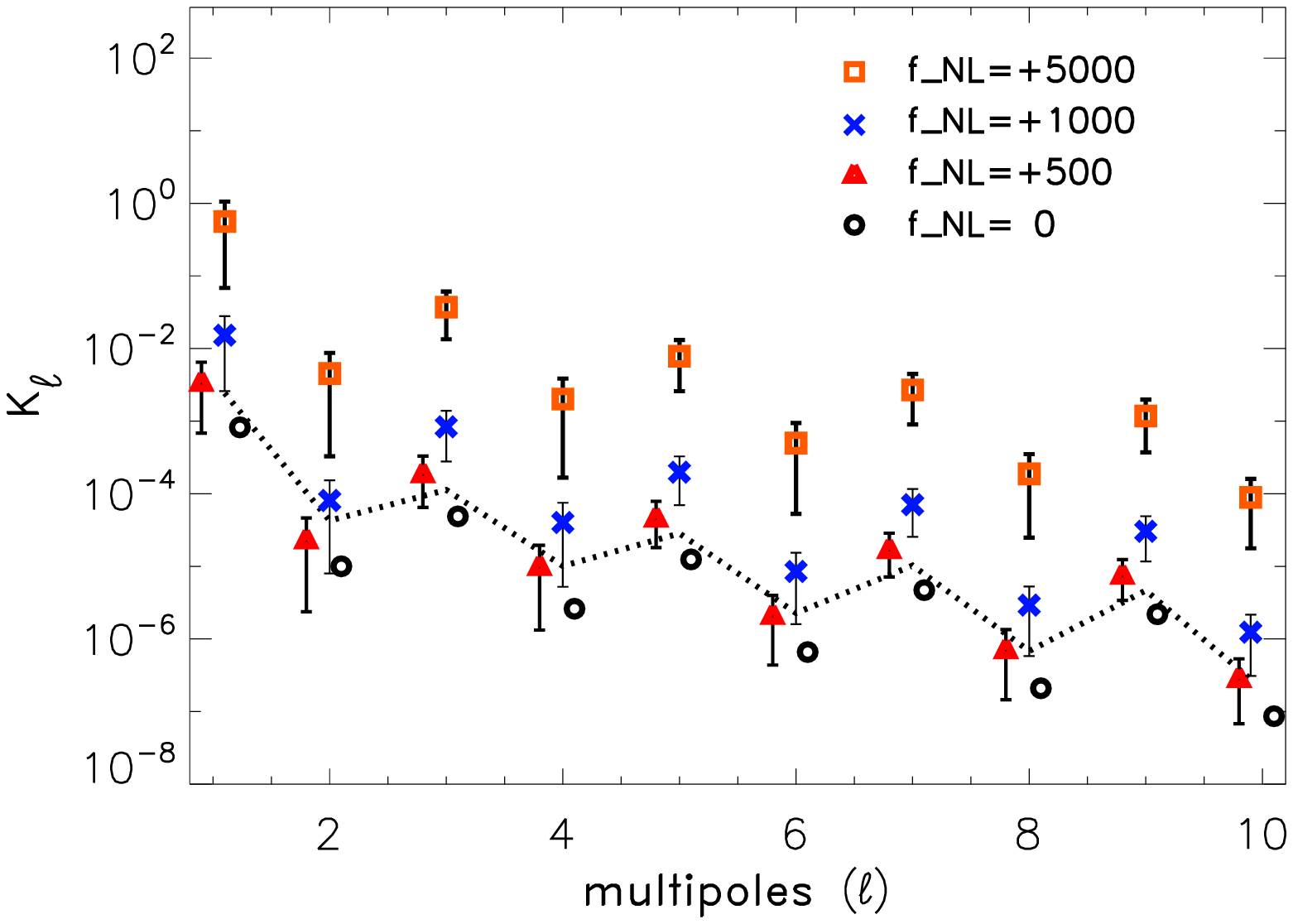}
\caption{Low $\ell$ average power spectra of skewness $S_{\ell}$ (left panel) and
kurtosis (right panel)
$K_{\ell}$ calculated from the Gaussian ($f_{\rm NL}^{\rm local}= 0$) and non-Gaussian
($f_{\rm NL}^{\rm local}= 500, 1\,000, \;\mbox{and} \;5\,000 $) input simulated CMB maps.
The $1\sigma$ error bars are also indicated with a small horizontal shift to avoid overlap.
The $95\%$ confidence level relative to the Gaussian maps is indicated by the dotted line.}
\label{Fig3}
\end{center}
\end{figure*}

\begin{figure*}[bht!]
\begin{center}
\includegraphics[width=8.8cm,height=5.6cm]{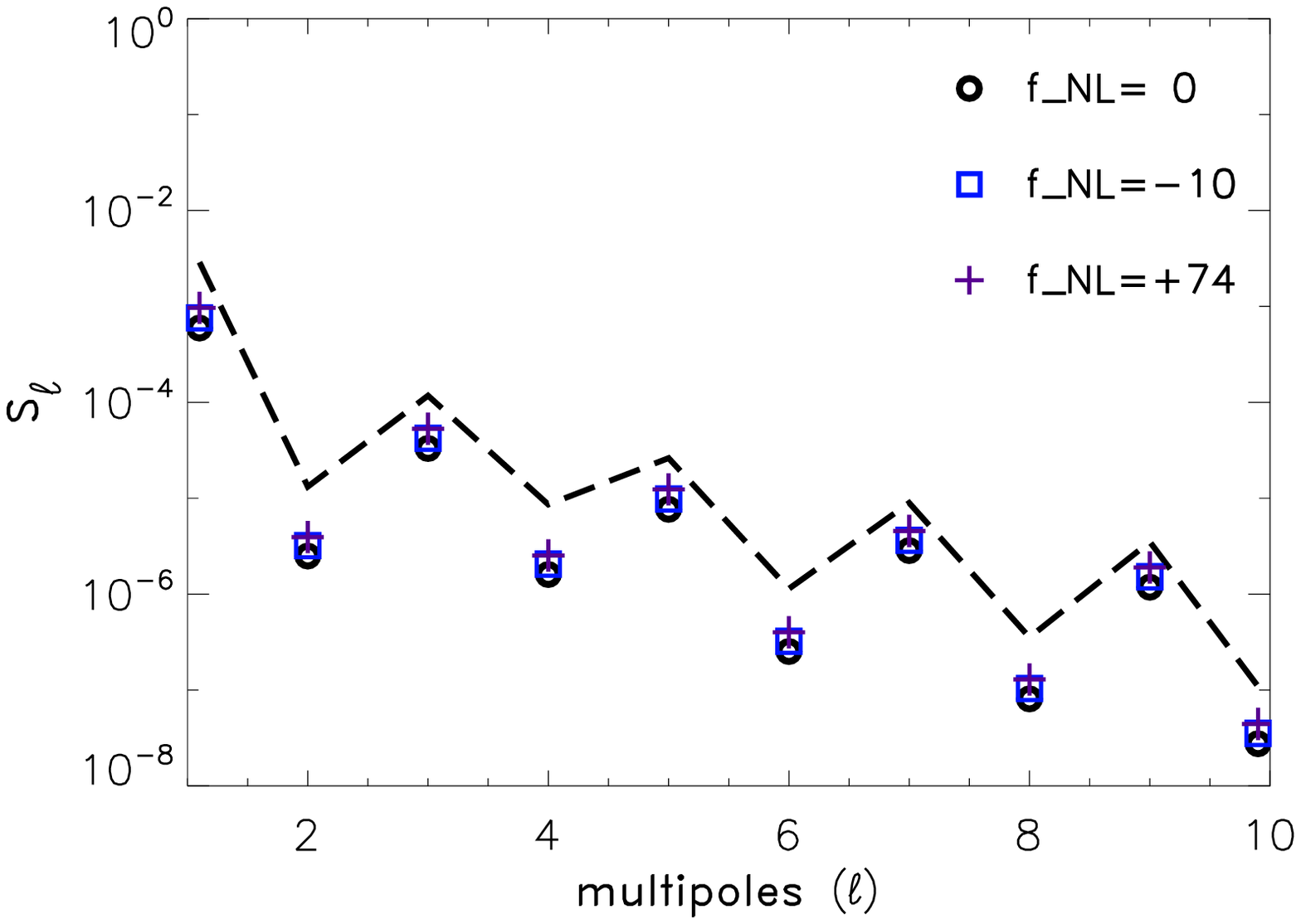}
\includegraphics[width=8.8cm,height=5.6cm]{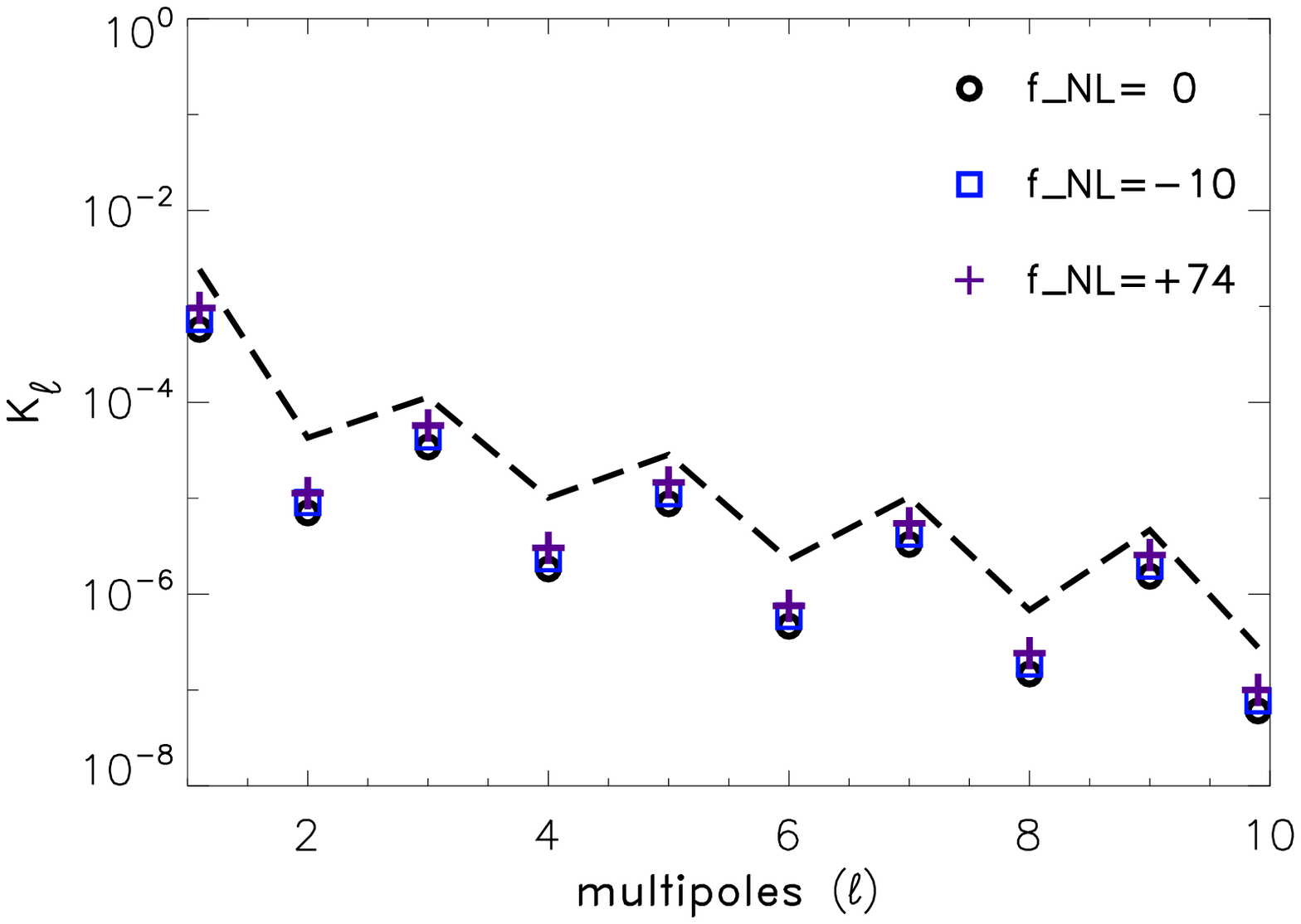}
\caption{Low $\ell$ average power spectra of skewness $S_{\ell}$ (left panel) and kurtosis
(right panel) $K_{\ell}$ calculated from the Gaussian ($f_{\rm NL}^{\rm local}= 0$) and
non-Gaussian simulated CMB maps for $f_{\rm NL}^{\rm local}= -10$ and
$f_{\rm NL}^{\rm local}= 74 $.
The $95\%$ confidence level relative to the Gaussian maps is indicated by the dashed line.}
\label{Fig4}
\end{center}
\end{figure*}

Figure~\ref{Fig3} shows the average power spectra of the skewness $S_{\ell}$
(left panel) and kurtosis $K_{\ell}$ (right panel), for $\,\ell=1,\,\,\cdots,10\,$,
calculated from simulated Gaussian ($f_{\rm NL}^{\rm local}= 0$) maps, and from
CMB maps equipped with non-Gaussianity of the local type for which
$f_{\rm NL}^{\rm local}= 500, \,1\,000, \,5\,000$.
The $95\%$ confidence level, obtained from the $S$ and $K$ maps calculated from
the Gaussian CMB simulated maps, is indicated in this figure by the dotted line.

To the extent that the average $S_{\ell}$ and $K_{\ell}$ obtained from input simulated
CMB maps endowed with $f_{\rm NL}^{\rm local}= 500$  
are within $95\%$ Monte-Carlo (MC) average values of $S_{\ell}$ and $K_{\ell}$
for $f_{\rm NL}^{\rm local}= 0$, Fig.~\ref{Fig3} shows that our indicators
are not suitable to detect  primordial non-Gaussianity of local type in CMB maps
smaller than this level. However, this figure also shows that they can
be effectively employed to detect higher levels of non-Gaussianity of local type.
These results square with the statistical analysis we shall report
in the remainder of this paper, particularly with the $\chi^2$ test of
goodness of fit.

To have an overall assessment power spectra $S_\ell$ and $K_\ell$,
calculated from the input simulated non-Gaussian maps equipped with
primordial non-Gaussianity with nonlinear parameter
$f_{\rm NL}^{\rm local}= 500, \,1\,000, \,5\,000$,
we have made a $\chi^2$ test to find out the goodness of
fit for  $S_{\ell}$ and $K_{\ell}$ multipole values as compared to
the MC mean multipole values obtained from $S$ and $K$ maps of the
Gaussian ($f_{\rm NL}^{\rm local}= 0$)  simulated maps.
In each case, this gives a number that quantifies collectively the
deviation from Gaussianity.
For the power spectra $S_\ell$ and  $K_\ell$ we found the values
given in Table~\ref{Tab:Chi2-Table1} for the ratio (reduced  $\chi^2$)
$\chi^2/\text{dof}\,$ (dof stands for degrees of freedom) for the power
spectra calculated from non-Gaussianity of local type with 
$f_{\rm NL}^{\rm local}= 500, \,1\,000, \,5\,000$.   
Moreover, the greater the reduced  $\chi^2$  values (hereafter denoted
simply by  $\chi^2$ ), the smaller the $\chi^2$ probabilities, that is,
the probability that the multipole values $S_{\ell}$ and $K_{\ell}$
and the mean multipole values obtained from the Gaussian maps
($f_{\rm NL}^{\rm local}=0$) agree.

\begin{table}[!bht]
\begin{center}
\begin{tabular}{ccc} 
\hline \hline 
Non-Gaussian Parameter \ \  &  \ $\chi^2_{}$ for $S_\ell$ \ \ \ \ & \ $\chi^2_{}$ for $K_\ell$ \ \\
\hline
$f_{\rm NL}^{\rm local}= 500$    & $  2.10                 $  & $ 2.12 \times 10 $ \\
$f_{\rm NL}^{\rm local}= 1\,000$   & $ 3.02  \times 10^{}  $  & $ \ 5.50 \times 10^2$  \\
$f_{\rm NL}^{\rm local}= 5\,000$   & $ 5.54  \times 10^3   $  & $ \ 1.31 \times 10^6 $  \\
\hline \hline
\end{tabular}
\end{center}
\caption{Results of the reduced $\chi^2$ test of the goodness of fit for $S_{\ell}$ and
$K_{\ell}$ calculated from the maps with different levels of non-Gaussianity as compared
with the corresponding mean values obtained from MC simulated CMB input maps with
$f_{\rm NL}^{\rm local}= 0$.}
\label{Tab:Chi2-Table1}
\end{table}

Thus, Table~\ref{Tab:Chi2-Table1}, along with Fig.~\ref{Fig3}, shows that the statistical
indicators $S$ and $K$ can clearly detect deviation from Gaussianity of simulated
maps endowed with primordial non-Gaussianity of local type with
$f_{\rm NL}^{\rm local} \gtrsim 500$.

A question that arises at this point is whether $S$ and $K$ indicators have sufficient
sensitivity to detect deviation from non-Gaussianity for the values of the non-Gaussian
parameter within the WMAP 7 yr bounds $ -10 <  f_{\rm NL}^{\rm local} < 74$%
~\cite{WMAP7}.
Table~\ref{Tab:Chi2-Table2} makes clear that for maps with $f_{\rm NL}^{\rm local}$
within this interval, one has a negligible value of $\chi^2$, which makes it apparent that
there is no significant overall departure of power spectra $S_\ell$ and $K_\ell$
for $f_{\rm NL}^{\rm local}= -10$ and $f_{\rm NL}^{\rm local}=  74$ from the
corresponding MC mean power spectra obtained from the Gaussian
($f_{\rm NL}^{\rm local}= 0$) maps, in agreement with the nearly overlapping
symbols of  Fig.~\ref{Fig4}.
This makes it apparent that the bispectrum based estimator that was employed by the
WMAP team~\cite{Komatsu-etal-2003,WMAP7} is more sensitive to primordial
non-Gaussianity of local type than the $S$ and $K$ indicators used in the present paper
(see also the related references~\cite{Related}).
Thus, for example, the deviation from Gaussianity as captured by our indicators $S$ and
$K$ for simulated maps with $f_{\rm NL}^{\rm local}=  74$ is 4 orders of magnitude
smaller than that for maps with $f_{\rm NL}^{\rm local}= 1\,000$.

To close this section, some words of clarification regarding the forthcoming CMB data
from the Planck mission are in order.
Planck combines high-angular resolution and sensitivity with a wide frequency coverage
($20$ GHz to $1\,000$ GHz) and will allow greatly improved foreground removal,
thereby reducing many sources of non-Gaussian contaminants.
The latest constraint on $f_{\rm NL}^{\rm local} $ from the WMAP-7 yr is
$f_{\rm NL}^{\rm local} = 32 \pm 21$ ($68\%$ confidence level),
and from Planck we expect $\Delta f_{\rm NL}^{\rm local} \simeq 5$.
This includes cosmic variance and the detector noise~\cite{Komatsu-Spergel_2000}.
For a similar reduction at $2\sigma$ confidence level, for example, the non-Gaussian
procedure of our paper does not have enough sensibility to detect deviation from
Gaussianity within the resulting Planck range of $f_{\rm NL}^{\rm local}$.
A preliminary analysis shows that a new procedure with a different way of scanning
input maps (through large pixels cells) seems to be capable of detecting such a
tiny deviation from Gaussianity, i.e. within a narrower range of
$f_{\rm NL}^{\rm local}$ expected from the Planck mission.

\begin{table}[!bht]
\begin{center}
\begin{tabular}{cll} 
\hline \hline 
Non-Gaussian Parameter \ \  &  \ $\chi^2_{}$ for $S_\ell$ \ \ \ \ \ & \ $\chi^2_{}$ for $K_\ell$ \ \\
\hline
$f_{\rm NL}^{\rm local}= -10$    & $ 2.80 \times 10^{-5}$  & $ 4.80 \times 10^{-5} $   \\
$f_{\rm NL}^{\rm local}= +74$     & $ 1.10 \times 10^{-3}$  & $ 1.00  \times 10^{-2}$  \\
\hline \hline
\end{tabular}
\end{center}
\caption{Results of the $\chi^2$ test of the goodness of fit for $S_{\ell}$ and $K_{\ell}$
calculated from the maps with $f_{\rm NL}^{\rm local} = -10$ and $f_{\rm NL}^{\rm local}= 74$
relative to the corresponding mean values obtained from MC simulated CMB input maps
with $f_{\rm NL}^{\rm local}= 0$.}
\label{Tab:Chi2-Table2}
\end{table}

\begin{figure*}[thb!]
\begin{center}
\includegraphics[width=8.8cm,height=5.6cm]{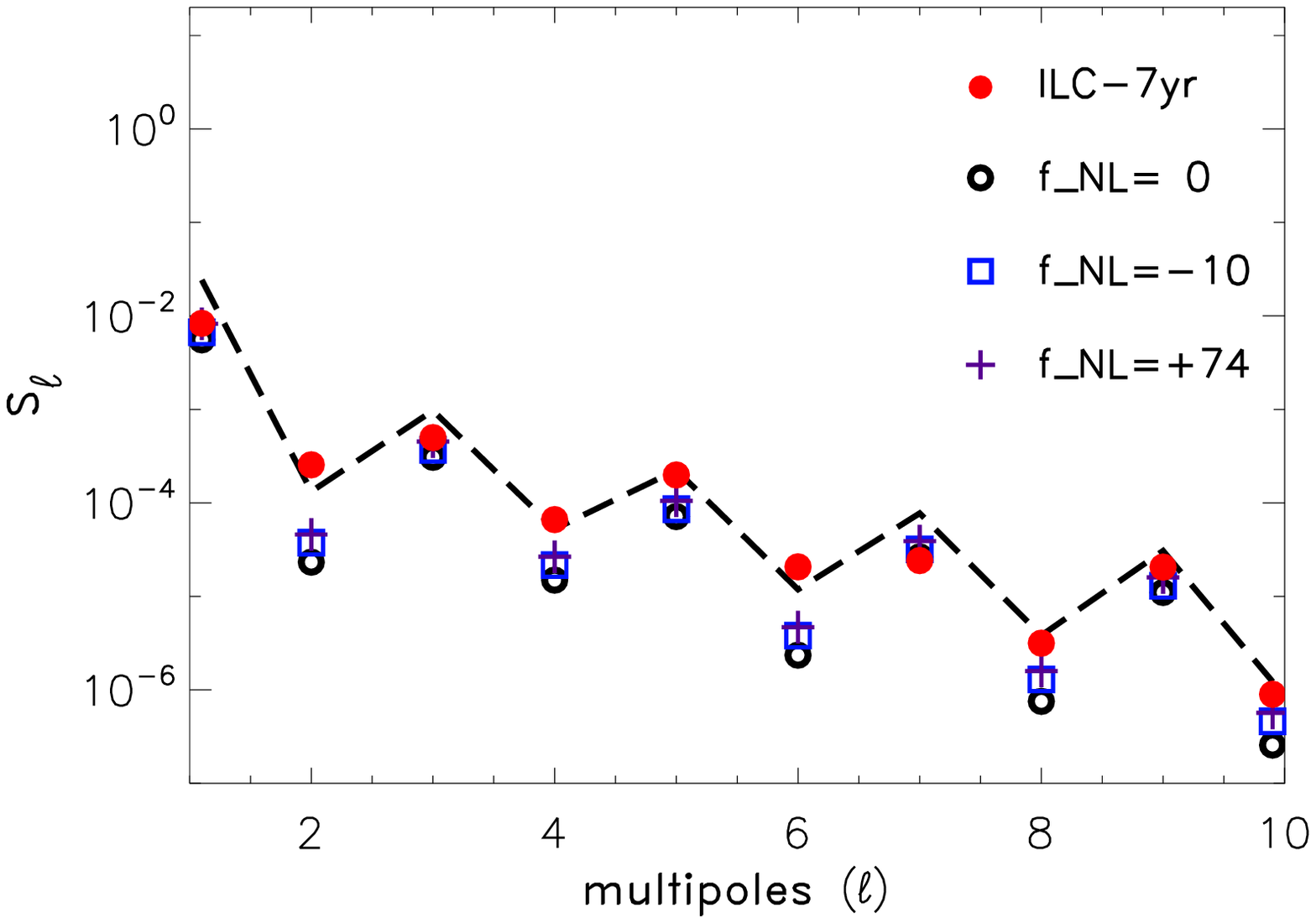}
\includegraphics[width=8.8cm,height=5.6cm]{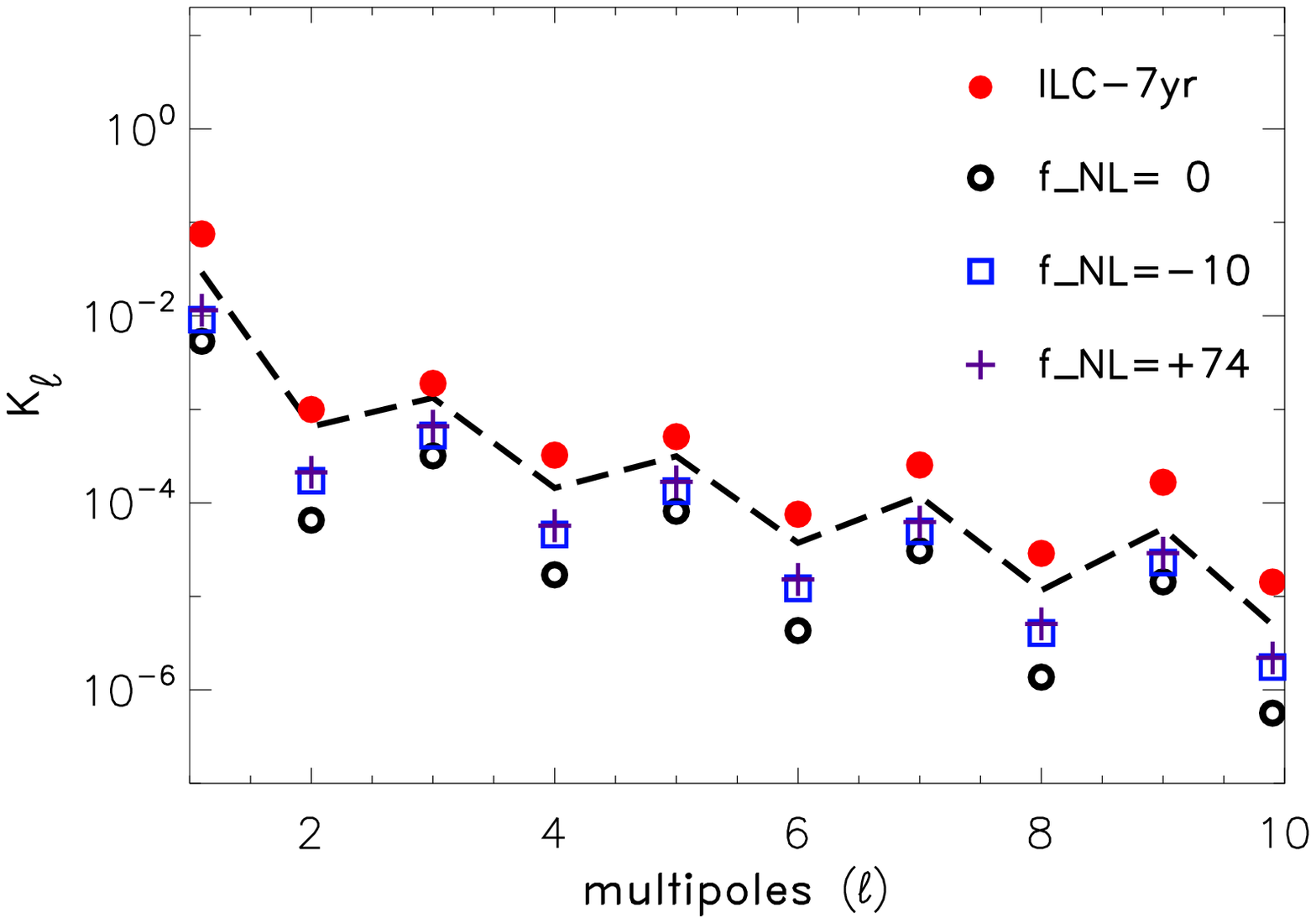}
\caption{Low $\ell$ mean power spectra of skewness $S_{\ell}$ (left panel) and
kurtosis (right panel) $K_{\ell}$ calculated from ILC-7 yr Gaussian
($f_{\rm NL}^{\rm local}= 0$) and non-Gaussian simulated CMB maps for
$f_{\rm NL}^{\rm local}= -10$ and $f_{\rm NL}^{\rm local}= 74 $.
In these calculations we have used simulated maps with the same smoothed $1^{\circ}$
resolution of the ILC-7 yr WMAP map.
The $95\%$ confidence level relative to the Gaussian maps is indicated by the dashed line.}
\label{Fig5}
\end{center}
\end{figure*}

\subsection{Analyses and results for simulated and data maps} 
\label{SubSec:Analy-Sim-data-Maps}

Besides the five frequency bands 7 yr maps released --K ($22.8$ GHz),
Ka ($33.0$ GHz), Q ($40.7$ GHz),  V ($60.8$ GHz), and W ($93.5$ GHz)--
the WMAP has also produced a full-sky foreground-reduced ILC-7 yr map, which is formed from a weighted linear
combination of these five frequency band maps in which the weights are
chosen in order to minimize the galactic foreground contribution.

The first-year ILC map has been explicitly stated as inappropriate for
CMB scientific studies~\cite{Bennett2003}, but in the subsequent ILC versions,
including the ILC-7 yr, a bias correction has been incorporated as
part of the foreground cleaning process, and the WMAP team suggested that these
maps are suitable for use in large angular scales (low $\ell$) analyses, although
they have not performed non-Gaussian tests on these versions of the ILC
maps~\cite{ILC-3yr-Hishaw,ILC-5yr-Hishaw,ILC-7 yr-Gold}.

In a recent paper~\cite{Bernui-Reboucas2010} we have performed an analysis of
Gaussianity of all the available  full-sky foreground-reduced maps by using
$S$ and $K$ non-Gaussianity indicators in the search for departure
from Gaussianity on large angular scales~\cite{Bernui-Reboucas2009}.
We have shown that the full-sky foreground-reduced WMAP maps, including
the ILC-7 yr maps~\cite{ILC-7yr-Gold}, present a significant deviation
from Gaussianity, in agreement with the results of
Ref.~\cite{Bernui-Reboucas2009,Bernui-Reboucas2010,Raeth-etal}.

An interesting question that arises here is how one compares the level of deviation
from Gaussianity of the full-sky ILC maps with the level of primordial non-Gaussianity
of local type with different amplitude parameters $f_{\rm NL}^{\rm local}$.%
\footnote{We note that as the simulated CMB maps are full-sky maps, and sky cuts
induce non-Gaussianity analyses. Thus, in order to make a comparative analysis
of Gaussianity between simulated and data maps, a simple suitable choice of CMB
data map is to take the equally full-sky data maps such as the ILC-7 yr.
Nevertheless, in Appendix~\ref{App_A} we present a comparative analysis between
maps with $f_{\rm NL}^{\rm local}$ for the WMAP-7 yr bounds and the
$Q$, $V$ and $W$ frequency maps.}
We  undertake this question by using our indicators
to carry out a comparative analysis of non-Gaussianity of simulated maps
for the values of non-Gaussian parameters  equal to the 
bounds reported by the WMAP team, and the ILC-7 yr map.

\begin{table}[!bht]
\begin{center}
\begin{tabular}{lcl} 
\hline \hline 
 Map   & \ \ \ \ $\chi^2_{}$ for $S_\ell$ \ \ \ \ & \ $\chi^2_{}$ for $K_\ell$ \ \\
\hline
$f_{\rm NL}^{\rm local}= -10$    & \ \ \ $  6.80 \times 10^{-2} $   & \ \ $ 3.80 \times 10^{-1} $   \\
$f_{\rm NL}^{\rm local}= +74$     & \ \ \ $  7.10  \times 10^{-2}$  & \ \ $ 4.24 \times 10^{-1} $  \\
ILC                              &  \ \ \ $ 8.96    $               & \ \ $ 5.88 \times 10      $  \\
\hline \hline
\end{tabular}
\end{center}
\caption{Results of the reduced $\chi^2$ test of the goodness of fit for $S_{\ell}$ and
$K_{\ell}$ calculated from ILC-7 yr and MC simulated maps for the 7 yr WMAP lower
and upper bounds values of $f_{\rm NL}^{\rm local}$. In these calculations we have
used simulated maps with the same smoothed $1^{\circ}$ resolution of the ILC-7 yr
WMAP map.
These values quantify collectively the deviation from Gaussianity relative to MC
Gaussian simulated maps ($f_{\rm NL}^{\rm local}=0$).}
\label{Tab:Chi2-Table3}
\end{table}

Figure~\ref{Fig5} shows the low $\ell$ mean power spectra of the skewness
$S_{\ell}$ (left panel) and kurtosis $K_{\ell}$ (right panel), calculated from
simulated Gaussian ($f_{\rm NL}^{\rm local}=0$) and non-Gaussian maps for
$f_{\rm NL}^{\rm local}= -10, 74$, which are the lower and upper bounds of the
nonlinear parameter reported recently by WMAP~\cite{WMAP7}.
We have calculated the power spectra by using $1\,000$ maps for each value of
$f_{\rm NL}^{\rm local}$.
This figure also displays $S_{\ell}$ and $K_{\ell}$ computed
from the full-sky foreground-reduced ILC-7 yr map.
Figure~\ref{Fig5} gives a clear indication that the deviation from Gaussianity of the ILC-7 yr
map (detected by $S$ and $K$) is greater than the non-Gaussianity of MC simulated maps
for the lower and upper bounds of $f_{\rm NL}^{\rm local}$ provided by
WMAP~\cite{WMAP7}.
Clearly, in the calculation of the mean power spectra $S_{\ell}$ $K_{\ell}$ of
Fig.~\ref{Fig5}, we have used simulated maps with the same smoothed $1^{\circ}$
resolution of the ILC-7 yr WMAP map.

In order to compare quantitatively  the deviation from Gaussianity, we have calculated
the $\chi^2$ test of the goodness of fit for $S_{\ell}$ and $K_{\ell}$ calculated from
ILC-7 yr along with the simulated maps for $f_{\rm NL}^{\rm local}= -10, 74$ relative to
the mean power spectra of Gaussian maps.
We have gathered together in Table~\ref{Tab:Chi2-Table3} 
the obtained values for the reduced $\chi^2$.
Figure~\ref{Fig5}, along with Table~\ref{Tab:Chi2-Table3}, shows, on the one hand,
a significant deviation from Gaussianity in the full-sky foreground-reduced ILC-7 yr
map, in agreement with the statistical analysis made in Ref.~\cite{Bernui-Reboucas2010}.
On the other hand, it is apparent that the level of non-Gaussianity of the ILC-7 yr maps
is higher than that of the simulated non-Gaussian maps with $f_{\rm NL}^{\rm local}$
equal to the bounds of the WMAP-7 yr data.
This comparison provides indications of the suitability of the foreground-reduced
ILC-7 yr map as Gaussian reconstructions of the CMB sky.

\section{Concluding remarks}

The physics of the  early  universe can be probed by
measurements of statistical properties of primordial fluctuations,
which are the seeds for the temperature CMB anisotropies.
Thus, the study of non-Gaussianity of these anisotropies
offers a powerful approach to probe the physics of the
primordial universe. It is essential, for example,  to
discriminate or even exclude classes of inflationary models,
and also to test alternative scenarios of the primordial universe.

Since one does not expect a single statistical estimator
to be sensitive to all possible forms of non-Gaussianity, it seems
important to test CMB data for deviations from Gaussianity by employing
different statistical tools to quantify or constrain the amount of any
non-Gaussian signals in the data, and extract information concerning
their potential origins.

In recent papers~\cite{Bernui-Reboucas2009,Bernui-Reboucas2010}
we proposed two new large-angle non-Gaussianity indicators, based on skewness
and kurtosis of large-angle patches of the CMB sky sphere and used them to find
significant large-angle deviation from Gaussianity in masked frequency bands and
foreground-reduced CMB maps.

Simulated CMB maps with an assigned primordial non-Gaussianity of a given type
and amplitude are important tools to study the sensitivity,  power, and limitations of
non-Gaussian estimators.
They can also be used to calibrate non-Gaussian statistical
indicators, and to study  the effects of foregrounds and
other non-Gaussian contaminants.

In this paper we have addressed the question as to whether the non-Gaussian
indicators proposed in Refs.~\cite{Bernui-Reboucas2009,Bernui-Reboucas2010}
have sufficient sensitivity to detect non-Gaussianity of local type, particularly with
amplitude $f_{\rm NL}^{\rm local}$ within the 7 yr WMAP bounds~\cite{WMAP7}.
To this end, we have used our statistical indicators
along with $6\,000$ simulated maps equipped with non-Gaussianity of local type with
various amplitudes.
From these simulated maps, which include the Gaussian one ($f_{\rm NL}^{\rm local}=0$),
we have generated $6\,000$ $S$ maps and $6\,000$ $K$ maps (see, e.g.,
Fig.~\ref{Fig1}), calculated  the associated low $\ell$ mean power spectra $S_\ell$ and
$K_\ell$, made a study of the sensitivity and strength, and determined the limitations of
non-Gaussian estimators $S$ and $K$.
By using the mean power spectra of the simulated non-Gaussian maps along
with the $\chi^2$ test of goodness, we have shown that $S$ and $K$ indicators
can be used to detect deviation from Gaussianity of local type for
typically $f_{\rm NL}^{\rm local} \gtrsim 500$ (see Fig.~\ref{Fig3} and
Table~\ref{Tab:Chi2-Table1}).
Thus, our indicators do not have enough sensitivity to detect deviation from Gaussianity
of local type with the nonlinear parameter within the 7 yr WMAP bounds.
This makes it apparent that the bispectrum based estimator employed by the
WMAP team is more sensitive to primordial non-Gaussianity of local type than
the $S$ and $K$.

However, by using the same procedure, which is based
upon the skeewness $S$ and kurtosis indicators $K$ of patches of the
CMB sky sphere, we have shown that these indicators do not have enough
sensitivity to capture non-Gaussianity
when the nonlinear parameter $f_{\rm NL}^{\rm local}$ lies
within the $95\%$ confidence level interval reported by the WMAP team~\cite{WMAP7}
(cf.\ Fig.~\ref{Fig4} and Table~\ref{Tab:Chi2-Table2}).
The positive outcome of the analysis performed in our previous works
\cite{Bernui-Reboucas2009,Bernui-Reboucas2010},
together with the present outcome, seems to indicate that the deviation from Gaussianity
captured in Refs.\cite{Bernui-Reboucas2009,Bernui-Reboucas2010}
is not of primordial nature, although it might have a primordial  
component.

Finally, we have also made a comparative study of non-Gaussianity of
simulated maps and of the WMAP full-sky foreground-reduced 7 yr ILC
map\cite{ILC-7yr-Gold}, which is summarized in Fig.~\ref{Fig5}
and Table~\ref{Tab:Chi2-Table3}.
An interesting outcome of this  analysis is that the level
of non-Gaussianity of ILC-7 yr ($f_{\rm NL}^{\rm local} \thicksim 775$) is
higher than that of the simulated maps for $f_{\rm NL}^{\rm local}$
within observational bounds\cite{WMAP7}.
This renders quantitative information about the  suitability of the
foreground-reduced ILC-7 yr map as a Gaussian reconstruction of the CMB sky.

\vspace{4mm}
\begin{acknowledgments}
M.J. Rebou\c{c}as acknowledges the support of FAPERJ under
a CNE E-26/101.556/2010 grant.
This work was also supported by Conselho Nacional de Desenvolvimento
Cient\'{\i}fico e Tecnol\'{o}gico (CNPq) - Brasil, under Grant No. 475262/2010-7.
A. Bernui thanks FAPEMIG for Grant APQ--01893--10.
M.J. Rebou\c{c}as and A. Bernui thank  CNPq for the grants under which this work
was carried out. We are grateful to A.F.F. Teixeira for reading the manuscript
and indicating some omissions and typos.
Some of the results in this paper were derived using the HEALPix
package\cite{Gorski-et-al-2005}.
We also acknowledge the use of the Legacy Archive for Microwave Background Data
Analysis (LAMBDA).
\end{acknowledgments}

\appendix
\section{}  \label{App_A}

Here we present the results of a comparative analysis of deviation from
Gaussianity performed by using Q, V, W band maps and simulated maps
equipped with non-Gaussianity of local type for
$f_{\rm NL}^{\rm local}= -10 \; \mbox{and} \; 74$,
which are the WMAP-7 yr bounds for $f_{\rm NL}^{\rm local}$.
As the calculations are similar to those of Sec.~\ref{Sec:Indic_vs_data},
we refer the readers to that section for more details on how they are made.

A word of clarification is in order before proceeding to the comparative
analysis. In order to have an estimation of the role of the masking process in
simulated maps with large values $f_{\rm NL}^{\rm local}$, we have calculated
the reduced $\chi^2$ of the mean $S_\ell$ and $K_\ell$ for maps endowed with
$f_{\rm NL}^{\rm local} = 500, 1\,000, 5\,000$ without and
with  the \emph{KQ75-7 yr} mask. These values quantify collectively the
deviation from Gaussianity relative to MC Gaussian simulated maps
($f_{\rm NL}^{\rm local}=0$).
Tables~\ref{Tab:Chi2-TableA1} and~\ref{Tab:Chi2-TableA2} contain the results of our
calculations.
These tables show that the greater $f_{\rm NL}^{\rm local}$ is the smaller the ratio is
between $\chi^2$  values calculated for masked and unmasked simulated maps,
respectively.
Hence, the relative role of non-Gaussianity induced by the \emph{KQ75-7 yr} mask in
simulated maps is smaller for higher values of $f_{\rm NL}^{\rm local}$, as we expected
from the outset.
Thus, the role of the masking process should be bigger for $f_{\rm NL}^{\rm local}$,
taking the lower and upper bounds values of the WMAP-7 yr, an issue that will be
discussed in the following.

\begin{table}[!bht]
\begin{center}
\begin{tabular}{ccl} 
\hline \hline 
Full-sky map & \ \ \ \ $\chi^2_{}$ for $S_\ell$ \ \ \ \ & \ $\chi^2_{}$ for $K_\ell$ \ \\
\hline
$f_{\rm NL}^{\rm local}= 500$     & \ \ \ $  2.10 $              & \ \ $  2.12 \times 10     $  \\
$f_{\rm NL}^{\rm local}= 1\,000$  & \ \ \ $  3.02 \times 10$     & \ \ $  5.50 \times 10^{2} $  \\
$f_{\rm NL}^{\rm local}= 5\,000$  & \ \ \ $  5.54 \times 10^{3}$ & \ \ $  1.31 \times 10^{6} $  \\
\hline \hline
\end{tabular}
\end{center}
\caption{Results of the reduced $\chi^2$ test of the goodness of fit for the mean power
spectra $S_{\ell}$ and $K_{\ell}$ as compared with the corresponding mean values
obtained from MC simulated CMB input maps with $f_{\rm NL}^{\rm local}= 0$.
Full-sky simulated maps were used.}
\label{Tab:Chi2-TableA1}
\end{table}

\begin{table}[!bht]
\begin{center}
\begin{tabular}{ccl} 
\hline \hline 
Masked map & \ \ \ \ $\chi^2_{}$ for $S_\ell$ \ \ \ \ & \ $\chi^2_{}$ for $K_\ell$ \ \\
\hline
$f_{\rm NL}^{\rm local}= 500$     & \ \ \ $  3.40 \times 10 $     & \ \ $  5.67 \times 10^{2} $  \\
$f_{\rm NL}^{\rm local}= 1\,000$  & \ \ \ $  4.69 \times 10 $     & \ \ $  6.81 \times 10^{2} $  \\
$f_{\rm NL}^{\rm local}= 5\,000$  & \ \ \ $  3.79 \times 10^{3}$  & \ \ $  4.79 \times 10^{5} $  \\
\hline \hline
\end{tabular}
\end{center}
\caption{Results of the reduced $\chi^2$ test of the goodness of fit for the mean power
spectra $S_{\ell}$ and $K_{\ell}$ as compared with the corresponding mean values
obtained from MC simulated CMB input maps with $f_{\rm NL}^{\rm local}= 0$.
The simulated input maps were masked with the \emph{KQ75-7 yr} mask.
\label{Tab:Chi2-TableA2}}
\end{table}

\begin{table}[!bht]
\begin{center}
\begin{tabular}{ccl} 
\hline \hline 
Full-sky map & \ \ \ \ $\chi^2_{}$ for $S_\ell$ \ \ \ \ & \ $\chi^2_{}$ for $K_\ell$ \ \\
\hline
$f_{\rm NL}^{\rm local}= -10$    & \ \ \ $  2.80 \times 10^{-5} $  & \ \ $  4.80 \times 10^{-5} $   \\
$f_{\rm NL}^{\rm local}= +74$   & \ \ \ $  1.10  \times 10^{-3}$  & \ \ $  1.00 \times 10^{-2} $  \\
Q               & \ \ \ $  8.79  \times 10^{14}$  & \ \ $ 1.75 \times 10^{23} $  \\
V               & \ \ \ $  6.15  \times 10^{14}$  & \ \ $ 3.17 \times 10^{23} $  \\
W              &  \ \ \ $ 3.14  \times 10^{13}$  & \ \ $ 9.14 \times 10^{21} $  \\
\hline \hline
\end{tabular}
\end{center}
\caption{Results of the reduced $\chi^2$ test of the goodness of fit for $S_{\ell}$
and $K_{\ell}$ calculated from Q, V, and W band maps and MC simulated maps for the
lower and upper bounds values of $f_{\rm NL}^{\rm local}$ obtained from WMAP-7 yr data.
These values quantify collectively the deviation from Gaussianity relative to MC
Gaussian simulated maps ($f_{\rm NL}^{\rm local}=0$).
Full-sky simulated and frequency maps were used.}
\label{Tab:Chi2-TableA3}
\end{table}

We begin the above-mentioned comparative analysis by recalling that one either has the
full-sky \emph{contaminated} (Q, V, W) band maps to compare with full-sky simulated maps,
or one masks both band and simulated maps.
While in the former we have foreground maps that are quite contaminated,
in the latter an induced non-Gaussianity arises from the masking process in simulated maps.

We have gathered together in Table~\ref{Tab:Chi2-TableA3} the values of the reduced
$\chi^2$ for the full-sky Q , V, and W frequency maps along with the values of the
reduced $\chi^2$ for maps endowed with non-Gaussianity for
$f_{\rm NL}^{\rm local}= -10 \; \mbox{and} \; 74$.
This table shows that the full-sky  Q , V, and W  band maps present a non-Gaussianity,
as captured by our indicators, of several orders of magnitude higher than those of
simulated maps for the lower and upper bounds values of $f_{\rm NL}^{\rm local}$
obtained from WMAP-7 yr data. Clearly, this is not a surprising result since the
full-sky Q, V, and W band maps are very foreground contaminated, but it shows
the suitability of our skewness and kurtosis indicators to detect such a huge
difference in the levels of non-Gaussianity.

We have collected together in Table~\ref{Tab:Chi2-TableA4} the results of the calculations
of the reduced $\chi^2$ for masked maps of the Q, V, and W bands, and for simulated input
maps with $f_{\rm NL}^{\rm local}= -10 \; \mbox{and} \; 74$. Tables~\ref{Tab:Chi2-TableA3}
and~\ref{Tab:Chi2-TableA4} show that, as expected, the \emph{KQ75-7 yr} mask reduced
significantly the levels of non-Gaussianity of bands maps, bringing them down several
orders of magnitude.
Even with the mask the detected level of non-Gaussianity for the Q band map is about
4 times the small level of the simulated masked maps for the WMAP-7 yr upper bound.
As for the  V and W bands and simulated maps, Table~\ref{Tab:Chi2-TableA4} shows that
these bands maps present deviation from Gaussianity, as measured by our indicators, of
similar order to those of maps with $f_{\rm NL}^{\rm local}$ in the WMAP-7 yr
interval.

\begin{table}[!bt] \vspace{6mm}
\begin{center}
\begin{tabular}{ccl} 
\hline \hline 
Masked map & \ \ \ \ $\chi^2_{}$ for $S_\ell$ \ \ \ \ & \ $\chi^2_{}$ for $K_\ell$ \ \\
\hline
$f_{\rm NL}^{\rm local}= -10$    & \ \ \ $0.79$   & \ \  $0.59$   \\
$f_{\rm NL}^{\rm local}= +74$   & \ \ \ $0.90$   & \ \  $0.85$  \\
Q                                               & \ \ \ $4.04$   & \ \ $3.20$  \\
V                                               & \ \ \  $0.86$   & \ \ $0.56$  \\
W                                              &  \ \ \ $0.73$   & \ \ $0.60$  \\
\hline \hline
\end{tabular}
\end{center}
\caption{Results of the reduced $\chi^2$ test of the goodness of fit for $S_{\ell}$
and $K_{\ell}$ calculated from Q, V, and W band maps and MC simulated maps
for WMAP-7 yr lower and upper bounds values of $f_{\rm NL}^{\rm local}$.
The \emph{KQ75-7 yr} mask was used for both frequency and simulated maps.
These values quantify collectively the deviation from Gaussianity relative to MC
Gaussian simulated maps ($f_{\rm NL}^{\rm local}=0$).}
\label{Tab:Chi2-TableA4}
\end{table}

\vspace{1.5cm}

\end{document}